\documentclass[%
 reprint,
superscriptaddress,
 amsmath,amssymb,
 aps,
orb,
]{revtex4-2}

\usepackage{graphicx}% Include figure files
\usepackage{dcolumn}% Align table columns on decimal point
\usepackage{bm}% bold math
\usepackage[colorlinks,linkcolor=blue,anchorcolor=blue,citecolor=blue,urlcolor=blue]{hyperref}% add hypertext capabilities
\usepackage{xcolor}
\usepackage[mathlines]{lineno}% Enable numbering of text and display math
\usepackage[textcolor=red,textsize=small,textwidth=0.6in,backgroundcolor=white,bordercolor=white]{todonotes}

\begin{document}

%\preprint{APS/123-QED}

\title{Metallic Contact Contributions in Thermal Hall Conductivity Measurements}% Force line breaks with \\

\author{Hongyu Ma}
 %Lines break automatically or can be forced with \\

\author{Xuesong Hu}
\affiliation{International Center for Quantum Materials, Peking University, Beijing 100871, China}

\author{Junren Shi} \email{junrenshi@pku.edu.cn}

\affiliation{International Center for Quantum Materials, Peking University, Beijing 100871, China}

\affiliation{Collaborative Innovation Center of Quantum Matter, Beijing 100871, China}

\date{\today}% It is always \today, today,
             %  but any date may be explicitly specified

\begin{abstract}
We investigate the influence of metallic contacts on thermal Hall measurements.
By analyzing typical measurement setups, we show that heat currents bypassing through metallic contacts could generate non-negligible thermal Hall signals.
% even when the measured insulator has zero thermal Hall conductivity.
We find that contributions from metallic contacts with thicknesses on the order of 10$^{-2}$ of sample widths can approximately replicate experimental observations across different materials in both temperature dependence and magnitude, assuming silver contacts with a conductivity of $10^{8}~\mathrm{S/m}$.
Our analysis underscores the need to minimize metallic contact effects in thermal Hall measurements, which can be achieved by optimizing measurement configurations.
%It even reproduces the subtle differences in temperature dependencies between materials with high and low longitudinal thermal conductivity.
\end{abstract}

%\keywords{Suggested keywords}%Use showkeys class option if keyword
                              %display desired
\maketitle

%\tableofcontents

\section{introduction}

Over the past two decades, the measurement of the thermal Hall effect (THE), as an effective means of probing charge-neutral heat carriers in insulators, has received much attention.
The effect was first detected in the paramagnetic insulator Tb$_3$Ga$_5$O$_{12}$\cite{PhysRevLett.95.155901,Inyushkin2007} and has since been measured in various other insulators, including cuprates~\cite{grissonnanche2019giant,grissonnanche2020chiral,boulanger2020thermal}, antiferromagnets~\cite{PhysRevB.108.L140402,chen2022large}, and multiferroic materials~\cite{ideue2017giant}.
% Several theoretical models have been proposed to explain the observed thermal Hall effect, considering both intrinsic and extrinsic mechanisms.
% The intrinsic mechanisms include phonon Berry curvature~\cite{PhysRevLett.105.225901,PhysRevB.86.104305,PhysRevB.85.134411,PhysRevLett.123.255901,hu2025intrinsic}, coupling between phonons and other quasiparticles such as magnons~\cite{PhysRevLett.96.155901,PhysRevLett.117.217205,PhysRevLett.123.167202,PhysRevB.104.035103}.
% Extrinsic mechanisms involve scattering of phonons by impurities or defects~\cite{PhysRevLett.124.167601,PhysRevB.105.L220301,PhysRevB.106.144111,PhysRevResearch.5.033197}.

Recently, many studies have reported the observation of ``giant'' THE~\cite{PhysRevX.10.041059,PhysRevLett.127.247202,PhysRevB.107.L060404,PhysRevResearch.5.043110,ataei2024phonon,PhysRevB.99.134419,kim2024thermal,PhysRevX.12.021025,PhysRevB.105.L220301,PhysRevLett.118.145902,nawwar2025large}.
It is generally believed that charge-neutral carriers like phonons only couple weakly with magnetic fields.
Therefore, the THE contributed by them is expected to be small.
However, in certain materials, such as cuprates~\cite{grissonnanche2019giant} and quantum paraelectrics~\cite{PhysRevLett.124.105901,PhysRevLett.126.015901}, measured thermal Hall conductivities are much larger than expected.
More surprisingly, giant thermal Hall conductivities have even been observed in some elemental substances or their simple compounds, such as black phosphorus, $\mathrm{Si}$ and $\mathrm{SiO_2}$~\cite{li2023phonon,jin2024discovery}. 
Theoretical efforts have been made to explain these observations.
In cuprates, the THE is usually attributed to phonon-spin interactions~\cite{grissonnanche2020chiral,boulanger2020thermal}, since spins can couple to magnetic fields directly.
For nonmagnetic insulators, studies primarily focus on extrinsic mechanisms involving phonons, such as phonon-domain wall scattering in SrTiO$_3$~\cite{PhysRevLett.124.167601}, since intrinsic mechanisms, like the phonon Berry curvature contributions from acoustic~\cite{PhysRevB.86.104305} and optical phonons~\cite{xy3k-mtrg}, yield predictions that are at least four orders of magnitude smaller than experimental observations.
There is still room for open inquiries into the origin of the effect.
%In cuprates, the THE may be attributed to phonon-spin interactions~\cite{grissonnanche2019giant,grissonnanche2020chiral}, since spins can couple to magnetic fields directly.
%For many other materials, such as BiSbTeSe$_2$\cite{PhysRevB.109.104304} and Y$_2$TiO$_7$\cite{PhysRevB.110.L100301}, there is still no plausible explanation for the occurrence of the giant THE.
 
A notable fact is that metallic components are commonly used in recent measurements of the THE.
In contrast, early experiments carefully avoided metallic components to minimize spurious contributions from conducting electrons~\cite{strohm2003magneto}.
% With the advancement of experimental setup, metallic components were introduced to ensure efficient thermal conduction\cite{inyushkin2007phonon,kim2019modular}. However, the impact of these metallic components on the experimental results remains unclear.
Later studies show that metallic heat sinks have negligible effects on measurements~\cite{Inyushkin2007,hirschberger2015large}, although there is a report suggesting otherwise~\cite{yamashita2019thermal}.
Besides heat sinks, metallic contacts are also widely employed to ensure efficient thermal connections~\cite{jiang2022phonon,chen2022large,PhysRevB.105.115101,PhysRevB.109.104304,uehara2022phonon,grissonnanche2020chiral}.

Employing metallic contacts in THE measurements introduces a potential issue.
When a heat current flows through a sample, a portion of the heat current will inevitably enter metallic contacts that are in direct contact with the sample.
Under a magnetic field, this bypass heat current can induce a transverse temperature gradient within the metallic contacts, giving rise to spurious thermal Hall signals.
Although these contacts are usually much smaller in geometry than the sample and may appear insignificant, their spurious contribution to the measured thermal Hall coefficients is not necessarily negligible, as metals have much larger thermal Hall conductivities than insulators.
This naturally raises a question: What is the influence of metallic contacts on the experimental results?

In this paper, we study the influence of metallic contacts on thermal Hall measurements.
By analyzing typical measurement configurations, we demonstrate that heat currents bypassing through metallic contacts can generate non-negligible thermal Hall signals, even when the measured material has zero thermal Hall conductivity.
We find that contributions from metallic contacts with thicknesses on the order of 10$^{-2}$ of sample widths can approximately replicate experimental observations across different materials in both temperature dependence and magnitude, assuming silver contacts with a conductivity of $10^{8}~\mathrm{S/m}$.
Our analysis underscores the need to minimize metallic contact effects in thermal Hall measurements.
We thus propose a measurement configuration to effectively suppress this influence. 
%It even reproduces the subtle differences in temperature dependencies between materials with high and low longitudinal thermal conductivity.

The remainder of the paper is organized as follows.
In Sec.~\ref{2}, we analyze the effect of metallic contacts and derive a formula for predicting the apparent thermal Hall conductivity of a sample with zero thermal Hall conductivity as measured using metallic contacts.
In Sec.~\ref{3}, we fit experimental data for various materials using the formula. The fits yield nominal contact thickness parameters, which can be used to assess the relative importance of the contact contributions.
Finally, we summarize our findings and propose a measurement configuration designed to suppress the influence of metallic components in Sec.~\ref{4}.

\section{\label{2}Metallic contact contributions}

In recent experiments, metallic contacts are widely employed in THE measurements.
Two types of contacts are commonly used.
The first type involves thin metallic electrodes attached to the sides of a sample, with contact lengths that are typically a fraction of the sample size~\cite{PhysRevLett.124.105901,jiang2022phonon}.
The second type consists of fine metallic wires, typically $25~\mu\mathrm{m}$ or $50~\mu\mathrm{m}$ in diameter, affixed to the sample surface~\cite{Inyushkin2007,hirschberger2015large,chen2022large,PhysRevB.105.115101,PhysRevB.109.104304,uehara2022phonon,grissonnanche2020chiral,kim2019modular,jin2024discovery}.
Silver and gold are the metals most commonly used,
and silver paint is often utilized for attaching contacts to a sample.
% They are then either connected to thermometers via metallic wires~\cite{PhysRevLett.124.105901} or directly bonded to the thermometers with silver paste~\cite{jiang2022phonon}.  One end is  using silver paste, and the other is connected to the thermometer~\cite{kim2019modular,jin2024discovery}.

\begin{figure}[htb]
\includegraphics[width=1\linewidth]{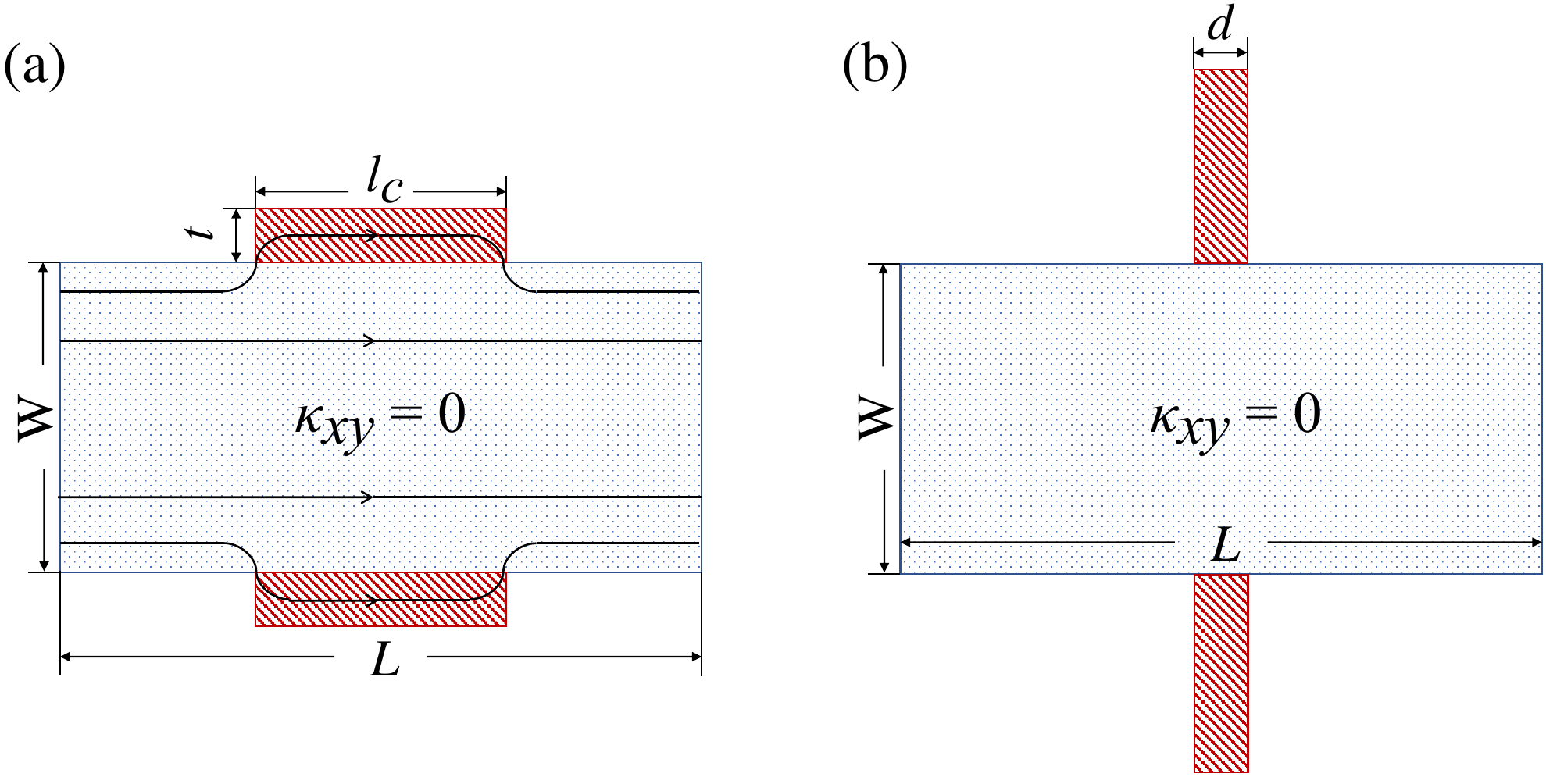}% Here is how to import EPS art
\caption{\label{fig.1} Simplified schematics of THE measurement configurations.
  (a) Electrode configuration:
  The red regions represent metallic electrodes with thickness $t$ and length $l_{c} \gg t$.
  The blue region represents the sample to be measured, with width $W$ and length $L$.
  Arrowed lines illustrate the heat current through the system and diverting into the electrodes.
  (b) Wire configuration:
  The red regions represent metallic wires with a diameter $d$.
  }
\end{figure}

For clarity in theoretical analysis, we consider two simplified configurations illustrated in Fig.~\ref{fig.1}.
The first configuration, referred to as the electrode configuration (Fig.~\ref{fig.1}(a)), consists of two thin metallic electrodes attached to the sides of the sample.
The second configuration, referred to as the wire configuration (Fig.~\ref{fig.1}(b)), consists of two fine metallic wires affixed to the sample.
For simplicity, we assume that the metallic contacts are of the same height as the sample, allowing us to analyze the configurations in two dimensions.
% set the electrical conductivity of the metal in contact with the material $\rho_{el}^\mathrm{c}=10^{-8}$, the Hall coefficient is taken to be that of silver,  -- it is too early to talk about this.
% We the thermal Hall conductivity of the insulator material $\kappa_{xy}^{\mathrm{m}}=0$.

We first analyze the electrode configuration.
When a heat current $J$ flows through the sample, a fraction of it enters the electrodes.
Since the electrode thickness $t$ is much smaller than its length $l_{c}$, the heat current through them is nearly uniform.
Therefore, the electrodes and sample can be treated as a parallel thermal circuit. The total bypass heat current through the two electrodes can be estimated by
\begin{equation}
\label{eq:J}
J_{c}\approx 2\frac{\kappa_{xx}^\mathrm{c}t}{\kappa_{xx}W+\kappa_{xx}^\mathrm{c}t}J\approx 2\frac{\kappa_{xx}^\mathrm{c}}{\kappa_{xx}}\frac{t}{W}J,
\end{equation}
where $\kappa_{xx}$ and $\kappa_{xx}^\mathrm{c}$ are the longitudinal thermal conductivities of the sample and the electrodes, respectively, and $W$ is the width of the sample. 
% In electrode and material, the heat current density are denoted as $j_{c}$ and $j $, with heat current $J_{c}=j_{c}t$ and $J=jW$. According to Ohm's law $j=\kappa \partial T$ and the parallel  formula, we obtain the relation $j_{c}/j=\kappa_{c}/\kappa $.  In the thermal Hall effect, the transverse temperature gradient is given by $\partial_y T=\rho_{xy}j_x-\rho_{xx}j_y$ where $j_x,j_y$ is the longitudinal and transverse heat current density,  $\rho_{xx},\rho_{xy}$ represents the thermal and thermal Hall resistivity. By integrating the transverse thermal gradient and assuming $j_y\ll j_x$,
The bypass heat current in the metallic electrodes induces a transverse temperature difference $\Delta T_y\approx\rho_{xy}^\mathrm{c}J_x$, where $\rho_{xy}^\mathrm{c}$ denotes the thermal Hall resistivity of the metallic electrodes.
In a THE measurement using this configuration, the transverse temperature difference could be misinterpreted as the result of a finite thermal Hall resistivity $\rho_{xy}^{\mathrm{app.}}=\Delta T_{y}/J$ even if the actual thermal Hall conductivity of the sample is zero.
It corresponds to an apparent thermal Hall conductivity:
\begin{equation}
\kappa_{xy}^{\mathrm{app.}}\approx\kappa_{xx}^{2}\rho_{xy}^{\mathrm{app.}}\approx2\rho_{xy}^\mathrm{c}\kappa_{xx}^\mathrm{c}\kappa_{xx}\frac{t}{W}.\label{eq.1}
\end{equation}

% Fig.~\ref{fig.1}(b) shows the wire configuration. In typical experimental setups, metallic wires with a diameter of approximately 25$\mu$m are commonly used. To simplify the geometric model, we assume that the wire is in direct contact with the material and has the same thickness as the material. Based on this assumption, the thermal Hall conductivity of the wires is derived (see Appendix~\ref{A}).

For the wire configuration, the heat current is expected to penetrate the wires over a distance comparable to their diameter $d$.
The approximate heat current distribution within the wires can be determined by solving Fourier's heat equation (see Appendix~\ref{A}).
The analysis indicates that a wire is equivalent to an electrode with an effective thickness
\begin{equation}
\label{eq:t}
t_{\mathrm{eff}} \approx 0.27 d.
\end{equation}
Eq.~\eqref{eq.1} can then be applied to determine the apparent thermal Hall conductivity.

% For simplicity, approximations were made in the preceding analyses.
A more realistic analysis based on finite-element simulations, presented in Appendix~\ref{B}, reveals that there exists a geometric correction that depends on the geometric parameter $t/l_{c}$ and the ratio $\kappa_{xx}/\kappa_{xx}^\mathrm{c}$.
With the correction, Eq.~(\ref{eq.1}) is modified as follows:
\begin{equation}
\kappa_{xy}^{\mathrm{app.}}=2\rho_{xy}^{{\mathrm c}}\kappa_{xx}^\mathrm{c}\kappa_{xx}\frac{t}{W}\delta\left(\frac{t}{l_{c}},\frac{\kappa_{xx}}{\kappa_{xx}^\mathrm{c}}\right),\label{eq:kmdelta}
\end{equation}
where $\delta(t/l_{c}, \kappa_{xx}/\kappa_{xx}^\mathrm{c})$ represents the correction factor, and can be approximated by the Pad\'{e} formula
\begin{equation}
\delta(p,q)=\frac{(\alpha+p)q}{\beta p+(\alpha+p)q},\label{eq:pade}
\end{equation}
with parameters $\alpha=0.355,\,\beta=0.694$.
The wire configuration corresponds to $t/l_{c} \rightarrow \infty$.
The correction factor could deviate significantly from unity when $\kappa_{xx}/\kappa_{xx}^\mathrm{c}$ is small.

\begin{figure*}[htb]
\includegraphics[width=1\linewidth]{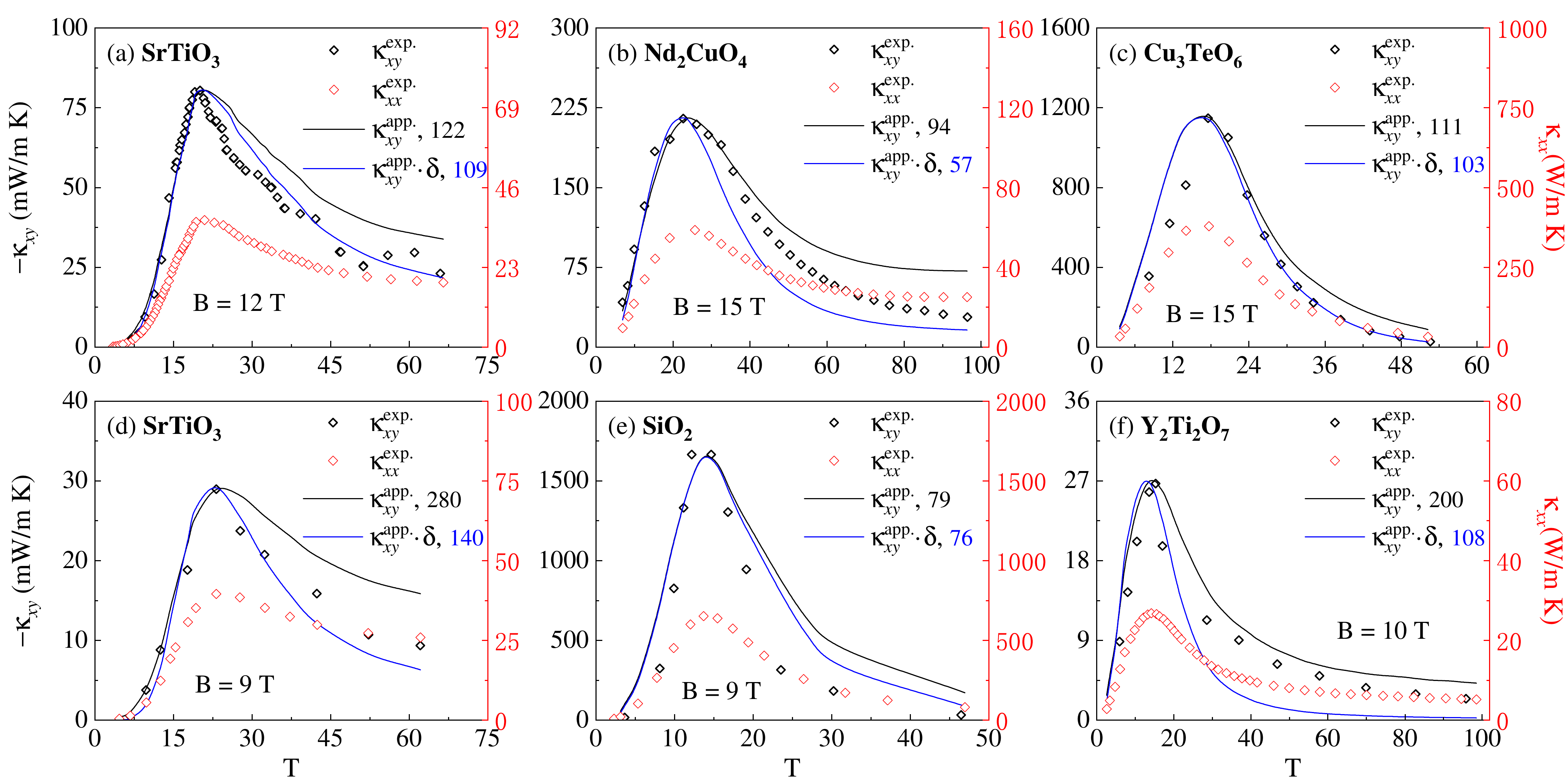}% Here is how to import EPS art
\caption{\label{fig.2} Fits to experimental thermal Hall conductivity ($\kappa_{xy}^{\mathrm{exp.}}$) for (a) SrTiO$_3$~\cite{PhysRevLett.124.105901}, (b) Nd$_2$CuO$_4$~\cite{boulanger2020thermal}, (c) Cu$_3$TeO$_6$~\cite{chen2022large}, (d) SrTiO$_3$~\cite{jin2024discovery}, (e) SiO$_2$~\cite{jin2024discovery} and (f) Y$_2$Ti$_2$O$_7$~\cite{PhysRevB.110.L100301}.
  Experimental longitudinal thermal conductivities ($\kappa_{xx}^{\mathrm{exp.}}$) reported alongside $\kappa_{xy}^{\mathrm{exp.}}$ are also shown. 
  Solid curves show fits to $\kappa_{xy}^{\mathrm{exp.}}$, with ($\kappa_{xy}^{\mathrm{app.}}\cdot \delta$) and without ($\kappa_{xy}^{\mathrm{app.}}$) the geometric correction, using Eq.~\eqref{eq:kmdelta} and Eq.~\eqref{eq.1}, respectively.
For geometric corrections, contact configurations specified in experiments are utilized: (a) uses the electrode configuration, assuming $t/l_{c}=1/20$, (b)--(f) use the wire configuration ($t/l_{c} = \infty$).
Numbers adjacent to the legends indicate the $W/t$ values used for fitting.
%Comparison of $\kappa_{xy}^{\mathrm{app.}}$ with $\kappa_{xy}^{\mathrm{exp.}}$ and $\kappa_{xx}^{\mathrm{exp.}}$ in different materials (a)~SrTiO$_3$\cite{PhysRevLett.124.105901}, (b)~Nd$_2$CuO$_4$\cite{boulanger2020thermal}, (c)~Cu$_3$TeO$6$\cite{10.1073/pnas.2208016119}, (d)~SrTiO$_3$\cite{jin2024discovery}, (e)~SiO$_2$\cite{jin2024discovery}, (f)~Y$_2$Ti$_2$O$_7$\cite{PhysRevB.110.L100301} and different configurations where the top row is electrode and the bottom row is wire. $\kappa_{xy}^{\mathrm{exp.}}$ and $\kappa_{xx}^{\mathrm{exp.}}$ are represented by black and red diamonds. The black and blue curves denote $\kappa_{xy}^{\mathrm{app.}}$ without and with the correction factor $\delta$. In the (a) and (b), $\delta$ is calculated using $t/d = 1/20$, while in the other panels $t/d \to \infty$ is used. The geometric parameter $t/W$ associated with each $\kappa_{xy}^{\mathrm{app.}}$ is labeled in the key, with the color matching the corresponding curve.
} 
\end{figure*}

We note that actual geometric configurations in experiments may differ in detail from the simplified configurations considered here.
Therefore, when applying the formula, the geometric parameters should be treated as effective ones, adjusted to the actual experimental setup.
For instance, in the electrode configurations, electrodes often cover a portion of the top and bottom surfaces of the sample.
Its effect can be approximated by scaling the thickness $t$ by a factor according to the increased total area of the electrode.
For more accurate estimates, one can always consider the actual contact geometry and employ finite-element simulations.
% \cite{jiang2022phonon,chen2022large,PhysRevB.105.115101,PhysRevB.109.104304,uehara2022phonon,grissonnanche2020chiral},
% they can all be effectively reduced to these configuration. Therefore, the thermal Hall  conductivity contributed by different configurations can be described by Eq. (\ref{eq.1}). In the following, we compare the results of Eq. (\ref{eq.1}) with the experimental results.

\section{\label{3}Fit to experimental results}

To evaluate the significance of metallic contact contributions in THE measurements, we fit experimental data from various materials obtained by different experimental groups using Eq.~(\ref{eq.1}) and Eq.~\eqref{eq:kmdelta} under the assumption that the actual thermal Hall conductivities of the samples are zero.
We adjust the $W/t$ parameter to match the peak value of $\kappa_{xy}^{\mathrm{app.}}$, determined using Eq.~(\ref{eq.1}) or Eq.~\eqref{eq:kmdelta}, with that of the experimental $\kappa_{xy}-T$ curve.
%We fit the experimentally measured $\kappa_{xy}$-$T$ relations simply by aligning their peak values with those of $\kappa_{xy}^{\mathrm{app.}}$ determined using Eq.~(\ref{eq.1}) or Eq.~\eqref{eq:kmdelta}.
Each fit yields a nominal value for the geometric parameter $W/t$.
By assessing whether the value falls within a realistic range, one can evaluate the relevance of metallic contact contributions in the specific THE measurement.

To facilitate the fitting, we need to assume a set of parameters, as their actual values vary across different experiments and are not always available.
The thermal conductivity $\kappa_{xy}^\mathrm{c}$ and the thermal Hall resistivity $\rho_{xy}^\mathrm{c}$ of the metallic contacts are derived from their electrical conductivity and Hall resistivity using the Wiedemann-Franz law~\cite{ashcroft2022solid}.
The electrical conductivity of the metallic contacts is set to $10^{8}~\mathrm{S/m}$, a value estimated for the composite structure comprising a silver electrode and a thin silver paint layer (see Appendix \ref{C}).
The temperature dependent electric Hall coefficient $R_{H}$ is taken from silver~\cite{smith1995low}.
For the longitudinal thermal conductivity of the sample, $\kappa_{xx}$, we use actual measurement data reported alongside thermal Hall results.

The fitted geometric parameter $t/W$ is nominal, as electrode parameters may not match actual experimental settings.
Additionally, thermal boundary resistance can significantly reduce $\kappa_{xx}^{\mathrm{c}}$ (see Appendix \ref{C}).
Consequently, $t/W$ should be scaled according to the actual parameters. 

\begin{figure}[htb]
\includegraphics[width=0.8\linewidth]{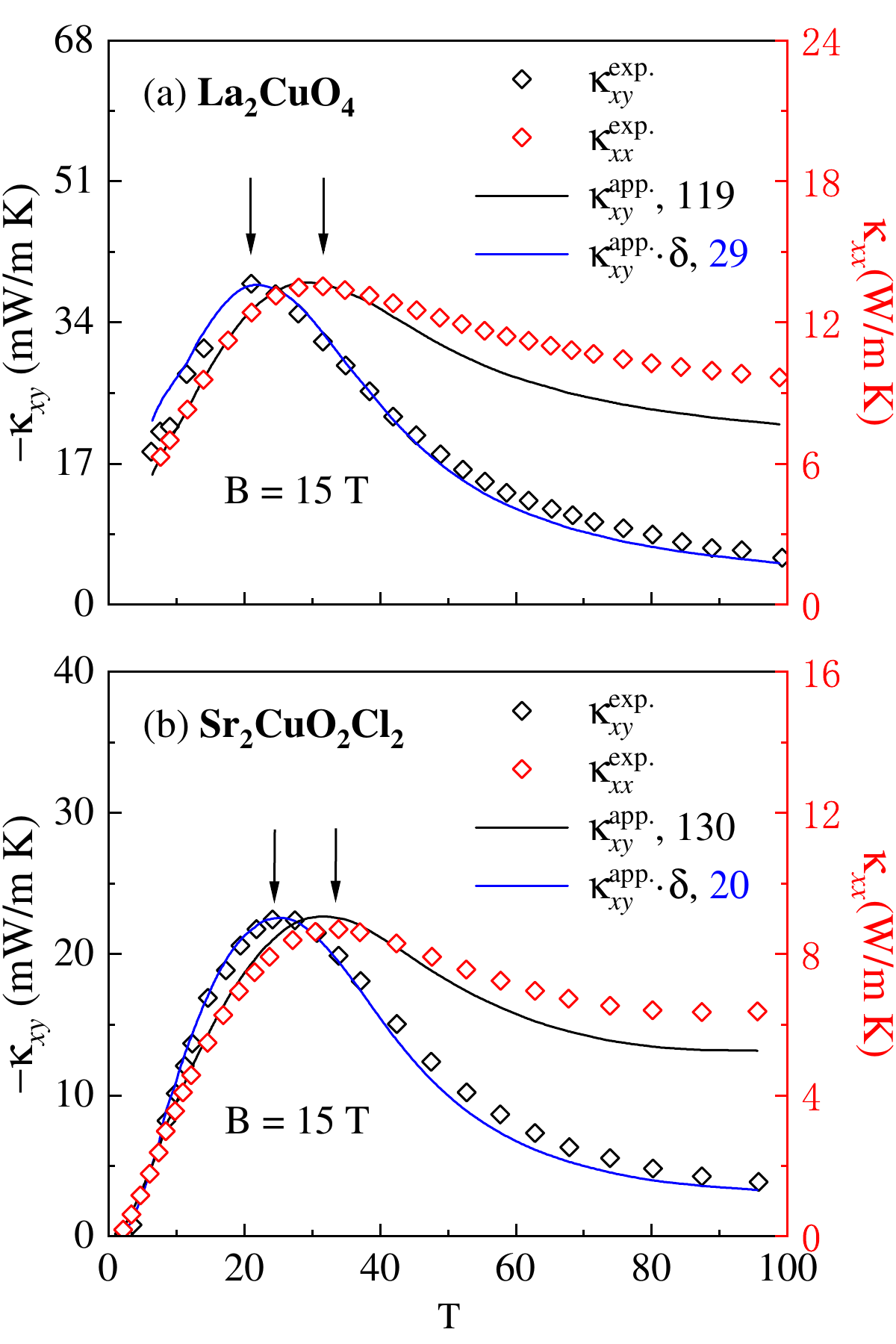}% Here is how to import EPS art
\caption{\label{fig.3} Same as Fig.~\ref{fig.2} but for (a) La$_2$CuO$_4$~\cite{boulanger2020thermal} and (b) Sr$_2$CuO$_2$Cl$_2$~\cite{boulanger2020thermal}.
  Arrows indicate the peak positions of $\kappa_{xx}^{\mathrm{exp.}}$ and $\kappa_{xy}^{\mathrm{exp.}}$.
  The wire configuration is assumed when determining geometric corrections.} 
\end{figure}

% Based on these physical parameters, the resulting $\kappa_{xy}^\mathrm{c}$ versus T curve is obtained.

Fits to a few representative experimental results from different studies are presented in Fig.~\ref{fig.2}.
The black and blue curves correspond to fits performed without and with the geometric correction, respectively.
The geometric parameter $W/t$ used for each fit is indicated in the legend.

It can be seen, above all, that the contact contribution $\kappa_{xy}^{\mathrm{app.}}$ reproduces the correct sign of the experimentally observed thermal Hall conductivity $\kappa_{xy}^{\mathrm{exp.}}$.
Experimental data exhibit a curious skewed distribution of $\kappa_{xy}^{\mathrm{exp.}}$ values, most of which are measured negative.
This happens to match the sign of the contact contribution, as the Hall coefficients of metals commonly used in contacts, such as Au and Ag, are negative.

Furthermore, $\kappa_{xy}^{\mathrm{app.}}$ exhibits temperature dependencies similar to those of $ \kappa_{xy}^{\mathrm{exp.}}$, with closely aligned peak temperatures.
Incorporating the geometric correction improves the quantitative agreement between $\kappa_{xy}^{\mathrm{app.}}$ and $\kappa_{xy}^{\mathrm{exp.}}$ in some cases.
According to Eq.~\eqref{eq.1}, the contact contribution is proportional to $\kappa_{xx}$,
implying that the peak temperature of $\kappa_{xy}^{\mathrm{app.}}$ coincides with that of $\kappa_{xx}$.
This is indeed observed in the experimental data shown in Fig.~\ref{fig.2}.

On the other hand, certain materials deviate from this behavior, exhibiting peak temperatures of $\kappa_{xy}^{\mathrm{exp.}}$ lower than those of $\kappa_{xx}$, as evident in Fig.~\ref{fig.3} for La$_{2}$CuO$_{4}$ and Sr$_{2}$CuO$_{2}$Cl$_{2}$.
Compared to the materials shown in Fig.~\ref{fig.2}, these materials have smaller $\kappa_{xx}$, falling in the regime where the geometric correction becomes significant. 
After incorporating the geometric correction, the contact contribution $\kappa_{xy}^{\mathrm{app.}}$ reproduces the observed shift of the peak temperatures, as evident in Fig.~\ref{fig.3}.

%This leftward shift is also reproduced in $\kappa_{xy}^{\mathrm{app.}}$.
%Analysis of Eq.~(\ref{eq:kmdelta}) indicates that the introduction of the correction factor $\delta$ leads to a shift of the $\kappa_{xy}^{\mathrm{app.}}$ peak toward lower temperatures.
%As the slope of $\delta$ increases, the magnitude of this shift becomes more pronounced.
%For the two materials in Fig.~\ref{fig.3}, $\delta$ varies rapidly near the $\kappa_{xx}$ peak temperature, resulting in a noticeable leftward shift of the $\kappa_{xy}^{\mathrm{app.}}$ peak.
%In contrast, $\delta$ for the materials in Fig.~\ref{fig.2} vary more gradually near the $\kappa_{xx}$ peak temperature, and thus no significant peak shift is observed in $\kappa_{xy}^{\mathrm{app.}}$. 

Most nominal $W/t$ values inferred from the fits are on the order of $10^{2}$.
However, incorporating geometric corrections in the two fits shown in Fig.~\ref{fig.3} reduces these values to as low as $20$.
Variations between experiments for the same material, such as SrTiO$_3$ in Fig.~\ref{fig.2}(a) and (d), could be attributed to differences in experimental configurations.
These values could serve as a basis for assessing the relevance of contact contributions in specific experiments.

\section{\label{4}Summary and Discussion}

In summary, we investigate the influence of metallic contacts on THE measurements.
Our analysis indicates that metallic contacts, despite having geometric dimensions much smaller than measured insulating samples, could induce a sizable thermal Hall signal.
Fits assuming THE arises solely from metallic contacts seem to reproduce the overall trends observed in experiments and account for sample-to-sample variations.

% This influence may cause an overestimation of the measurement, making it crucial to account for and exclude the metallic contact contribution to ensure the accuracy of the experimental results. 

% Metallic contacts are widely used in recent measurements of the THE, yet their influence on the measurement is often neglected.
% Therefore, it is necessary to consider and eliminate the spurious contribution of metallic contacts in experimental measurements.

%Metallic contacts are commonly present in thermal Hall effect experiments, yet their influence on the measurement is often overlooked. Notably, various configurations of metallic contacts can be systematically mapped onto equivalent configurations, providing a unified framework for evaluating their contribution to the thermal Hall effect. Our calculations show that metallic contacts can induce a significant transverse thermal conductivity, $\kappa_{xy}^\mathrm{c}$, over a realistic range of geometric dimensions, with peak overlapping with those of both $\kappa_{xx}^{\mathrm{ex}}$ and $\kappa_{xy}^{\mathrm{ex}}$. In certain cases, the electrode contribution alone is sufficient to reproduce the overall trend observed in experiments. These results highlight the necessity of carefully considering and, where possible, minimizing the role of metallic contacts when interpreting thermal transport measurements.

\begin{figure}[tb]
\includegraphics[width=1\columnwidth]{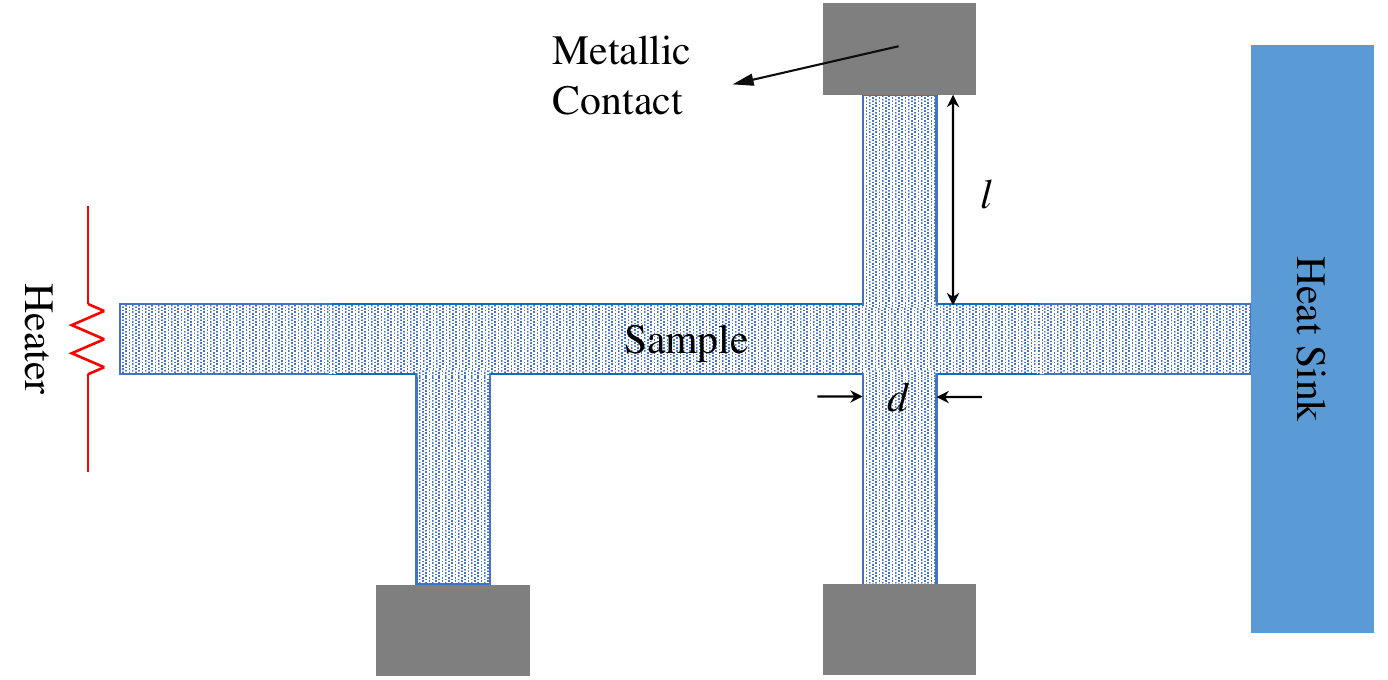}% Here is how to import EPS art
\caption{\label{fig.4} A sample shape designed to minimize the influence of metallic components in THE measurements.}
\end{figure}

Besides measurement contacts, metallic components in heat sinks and heaters could also adversely affect THE measurements.
Ref.~\cite{yamashita2019thermal} reports large spurious signals in a heat sink configuration with metallic components, though the effect is found negligible in other setups~\cite{Inyushkin2007,hirschberger2015large}. 
Moreover, a magnetic field can generate a temperature gradient within a heater, inducing a temperature difference in measurement contact, which could be wrongly interpreted as a thermal Hall conductivity.
% While estimating the magnitude of the spurious contributions is infeasible due to their dependence on the internal structures of heaters, they are expected to be proportional to $\kappa_{xx}^2$ rather than $\kappa_{xx}$.

% It is worth noting that there exist experimental observations that cannot be explained by the metallic contact contribution.
% For instance, the anisotropic compound investigated in Ref.~\onlinecite{grissonnanche2020chiral} exhibits a null THE signal in one direction but a finite signal in another. 
% Furthermore, an analysis in a recent study~\cite{xiang2025phonon} indicates that $\kappa_{xy}$ varies significantly more than $\kappa_{xx}$ during annealing cycles, violating the linear scaling relation expected by Eq.~\eqref{eq.1}.

To ensure the accuracy of THE measurements, it is crucial to minimize the influence of metallic components.
This can be achieved by using a sample-shape design borrowed from electrical Hall measurements~\cite{PhysRevLett.48.1559,4315329} and illustrated in Figure~\ref{fig.4}. 
In this design, metallic contacts, the heater, and the heat sink connect to the long probe arms of a sample.
In electrical measurements, long arms suppress current leakage into metallic probes~\cite{philips1958method,gluschke2020impact}.
The same principle applies to thermal measurements.
Moreover, the long arms connecting the heater and heat sink suppress the effects of transverse temperature gradients generated within these components.
As analyzed in Appendix \ref{B}, a perturbation at the end of a long arm (wire) is suppressed exponentially by a factor $\sim\exp(-\pi l / d)$, where $l$ and $d$ denote the length and width of the arm, respectively.
A probe arm with $l \approx 3d$ can achieve a suppression factor of $10^{-4}$.

%To ensure the accuracy of thermal Hall effect measurements, it is crucial to consider and eliminate the influence of metallic contacts. A practical solution is to minimize the use of metals in the experimental setup. Alternatively, heat current into metallic components can be suppressed. As discussed in Section II, the wire configuration reveals that the heat current density in the wire, $j_x=\sum_n4jl^2/(n^3\pi^3\kappa_{xx}^{\mathrm{m}})\exp(n\pi y/l)\sin(n\pi x/l)$, decays exponentially.  When the metal electrodes and wires are placed sufficiently far from the material, the amount of heat current entering these components becomes negligible. To achieve this, an insulating layer with adequate thickness must be introduced between the metal electrodes and the material. Under such conditions, the contribution of metal electrodes and wires to the thermal Hall signal can be effectively eliminated.

\begin{acknowledgments}
We acknowledge Zengwei Zhu, Shiyan Li, Jian Wang, Xi Lin, Gaël Grissonnanche and Yamashita Minoru for valuable discussions. This work is supported by the National Key R\&D Program of China under Grant No. 2021YFA1401900 and the National Science Foundation of China under Grant No. 12174005.
\end{acknowledgments}

\appendix

\section{\label{A}Heat current distribution in a wire}

To determine the bypass heat current entering a metallic wire connected to the sample, we consider an infinite wire extending along the $y$-axis from the origin to infinity, and occupying $x\in[0,d]$.
The temperature distribution can be determined by solving Fourier's heat equation $\nabla^2T=0$.
Assuming adiabatic boundary conditions at the lateral edges of the wire, the general solution takes the form
\begin{equation}
\label{eq.A1}
T(x,y)=T_0+\sum_n A_n e^{-\frac{n\pi}{d}y}\cos\left(\frac{n\pi}{d}x\right),
\end{equation}
where $A_n$ are coefficients to be determined.
At $y=0$, we assume that the temperature varies linearly, with a gradient matching that in the bulk of the sample: $T(x,y=0)=T_{0}+[J/(\kappa_{xx}W)]x$. 
Applying the boundary condition gives:
\begin{equation}
A_n=-\frac{4d}{n^2\pi^2\kappa_{xx}W}J.
\end{equation}

The bypass heat current density can be obtained using Fourier's law $j_x=-\kappa_{xx}^\mathrm{c}\partial_x T$.
The total bypass heat current along the $x$-direction, averaged over constant $x$ cross-sections, is given by
\begin{equation}
J_c= 2\sum_{n\in \mathrm{odd}}\frac{8\kappa_{xx}^\mathrm{c}d}{n^3\pi^3\kappa_{xx}W}J= 2\frac{\kappa_{xx}^\mathrm{c}}{\kappa_{xx}}\frac{7\zeta(3)}{\pi^3}\frac{d}{W}J\label{eq.A2}
\end{equation}
where the factor of $2$ accounts for two wires connected to the sample, and $\zeta(x)$ denotes the Riemann zeta function.
By comparing Eq.~(\ref{eq.A2}) with Eq.~(\ref{eq:J}) and noting $7\zeta(3)/\pi^{3}\approx 0.27$, we deduce the effective thickness for wires with a diameter $d$, as specified in Eq.~\eqref{eq:t}.

\begin{figure}[tb]
\includegraphics[width=\columnwidth]{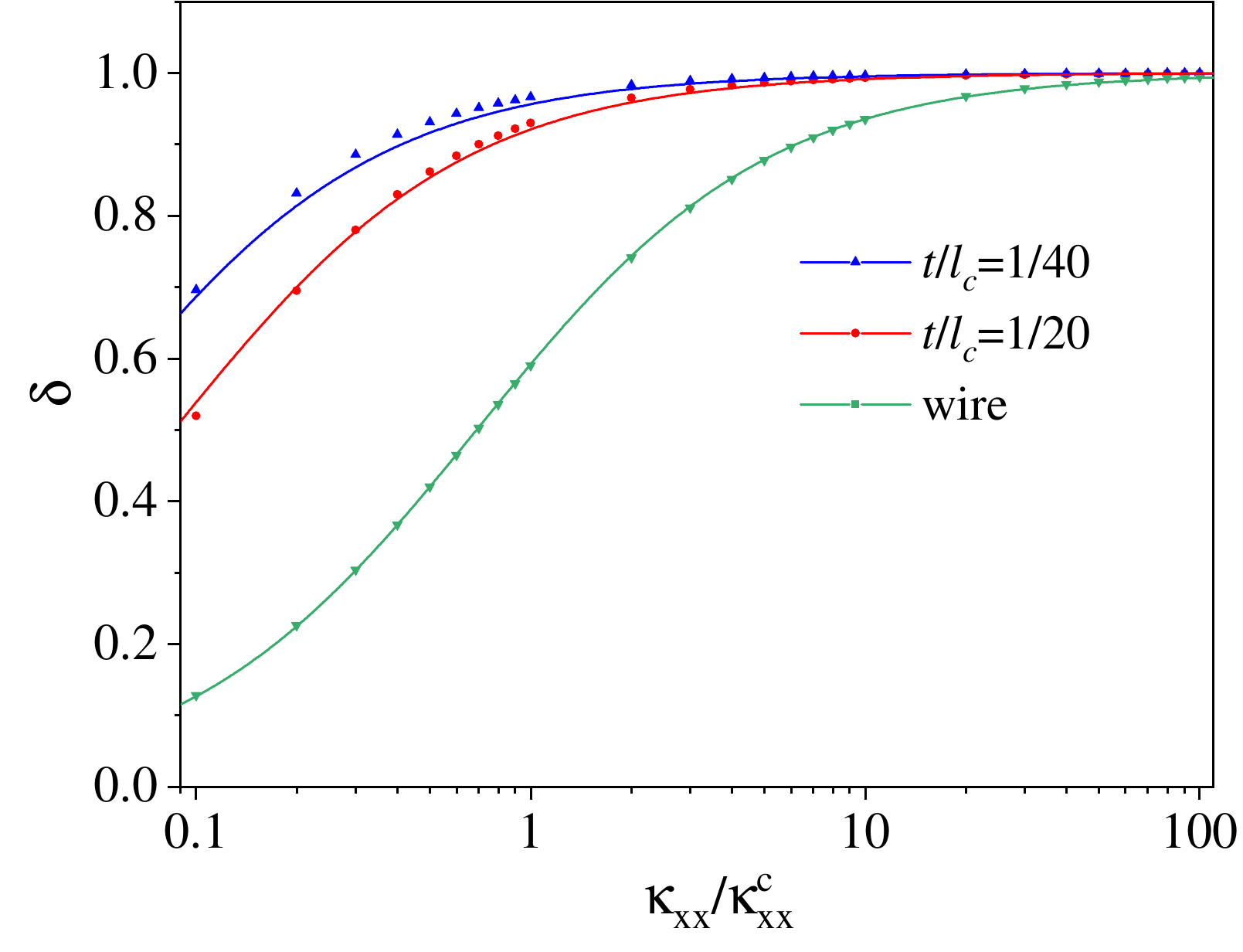}% Here is how to import EPS art
\caption{\label{fig.5} Geometric correction factor $\delta$ as a function of $\kappa_{xx}/\kappa_{xx}^\mathrm{c}$ and geometric ratio $t/l_{c}$. Points represent numerical results, while curves show values yielded by the Pad\'{e} formula Eq.~\eqref{eq:pade}. The wire configuration corresponds to $t/l_{c} \rightarrow \infty$.}
\end{figure}

% where $\kappa_{xx}^\mathrm{c}$ is the longitudinal thermal conductivities of the wires, and $\zeta$ is Riemann zeta function. Similar to the derivation of Eq.~(\ref{eq.1}), the measured thermal Hall conductivity would be:
% \begin{equation}
% \kappa_{xy}^{\mathrm{meas}}\approx2\rho_{xy}^{\mathrm c} \kappa_{xx}^{\mathrm c} \kappa_{xx}\frac{7\zeta(3)}{\pi^3}\frac{d}{W}\approx2\rho_{xy}^\mathrm{c}\kappa_{xx}^\mathrm{c}\kappa_{xx}\frac{0.27d}{W}
% \end{equation}

% Furthermore, when the wire is sufficiently short (with length $l$), Eq.~(\ref{eq.A3}) can be rewritten as $J_x = -\frac{2}{d}\int_0^d \int_0^{l}\kappa_{xx}^{
% \mathrm{c}
% }\partial_xT\mathrm{d}x
% \mathrm{d}y$. Expanding $T$ to first order in $y$ and performing the integration, we obtain $J_x = \kappa_{xx}^\mathrm{c}lJ/\kappa_{xx}W$.
% In this case, the measured thermal Hall conductivity $\kappa_{xy}^{\mathrm{meas}}\approx\rho_{xy}^{\mathrm c} \kappa_{xx}^{\mathrm c} \kappa_{xx} l/W$,
% which takes the same form as Eq.~(\ref{eq.1}). This result indicates that when the wire is sufficiently short, it can be regarded as an electrode, with the wire length $l$ corresponding to the electrode thickness $t$.

\section{\label{B}Geometric correction}

To evaluate the geometric corrections to Eq.~(\ref{eq.1}), we perform finite element simulations using the heat transfer module in COMSOL for the configurations shown in Fig.~\ref{fig.1}.
For all simulations, we fix $L/W = 2$ and $l_{c}/W = 0.4$.
Our calculations indicate that the correction factor $\delta$ is nearly independent of $l_{c}/W$ when $l_{c}/W \lesssim 1$.
For the electrode configuration, we vary $t/l_{c}$ between $1/20$ and $1/40$.
For the wire configuration, we set $t/l_{c} = 50$. A fixed temperature difference is applied across the two ends of the sample along its length, while all other boundaries are treated as adiabatic.
Grid density is adjusted to ensure numerical convergence.
A typical simulation employs approximately $2 \times 10^6$ grid points, with around $10^4$ allocated to the metallic contact.
In the electrode configuration, the total heat current passing through the center cross-section of the electrodes was computed.
For the wire configuration, we evaluated the average heat current in the wires.
The geometric correction factor $\delta$ is determined as the ratio of the numerical result to the approximate value yielded by Eq.~\eqref{eq:J}. 
% The simulations are based on Fourier's heat equation $\nabla^2 T = 0$.
% The longitudinal thermal conductivity of the metallic contact is set to 62.5616 $\mathrm{W/(m \cdot K)}$, and its thermal Hall resistivity is set to 0.0018 $\mathrm{m \cdot K/W}$.
% The longitudinal thermal conductivity of the sample is taken as a multiple of that of the metallic contact.

% The values of $\delta$ are shown in Fig.~\ref{fig.3}.

%Using the relations $\Delta T_y \approx \rho_{xy}^\mathrm{c}J_x$ and $\kappa_{xy} = -\kappa_{xx}\Delta T_yL/(\Delta T_xW)$, the numerical value of $\kappa_{xy}$ is obtained\todo{unnecessary}.

Figure~\ref{fig.5} shows the dependence of the correction factor $\delta$ on $\kappa_{xx}/\kappa_{xx}^\mathrm{c}$ for various contact geometries.
Significant geometric corrections are observed when $\kappa_{xx}/\kappa_{xx}^\mathrm{c} \lesssim 1$.
The correction is well approximated by the Pad\'{e} formula Eq.~\eqref{eq:pade}. 

\section{\label{C} Influences of silver paint and thermal boundary resistance}

\begin{figure}[htb]
\includegraphics[width=1\linewidth]{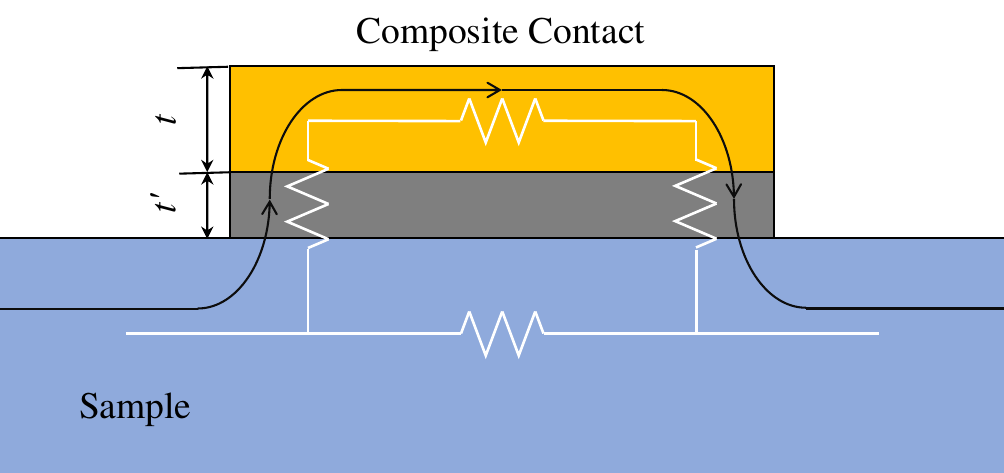}% Here is how to import EPS art
\caption{\label{fig.6} Schematic of heat current flow through a composite contact.
The yellow region represents electrodes with thickness $t$ and contact length $l_{c}$,
while the gray region represents a silver paint layer with thickness $t'$.
Arrowed lines indicate heat current flow, which is nearly perpendicular to the silver paint layer.
The equivalent series thermal circuit is also shown.
  }
\end{figure}

Metallic contacts are composite structures consisting of the silver paint and the metallic electrode (or wire). 
The effective thermal conductivity $\kappa^\mathrm{c}_{xx}$ is determined by the thermal conductivity of the silver paint $\kappa^{\mathrm{p}}_{xx}$ and that of the electrode $\kappa^{\mathrm{e}}_{xx}$. 
Since the silver paint has a much smaller thickness $t'$ than the electrode thickness $t$, the heat current approximately flows perpendicularly into and out of the silver paint layer, as shown in Fig.~\ref{fig.6}.  
In this case, the silver paint and the metallic electrode effectively form a series thermal circuit. 
The total thermal resistance of the composite structure can be approximated as
\begin{equation}
\frac{l_{c}}{\kappa^\mathrm{c}_{xx}(t+t')}\approx \frac{l_{c}}{\kappa^{\mathrm{e}}_{xx}t} + \frac{2t'}{\kappa^{\mathrm{p}}_{xx}(l_{c}/2)},
\end{equation}
which leads to
\begin{equation}
\frac{1}{\kappa^\mathrm{c}_{xx}} \approx \frac{1}{\kappa^{\mathrm{e}}_{xx}} \frac{t+t'}{t} + \frac{1}{\kappa^{\mathrm{p}}_{xx}}\frac{4 t'(t+t')}{l_{c}^{2}}\label{eq.C2}.
\end{equation}
% Eq.~(\ref{eq.C2}) provides a rough estimate of $\kappa_{xx}^\mathrm{c}$.
% A more refined evaluation would require introducing an additional factor that depends on multiple variables, making it difficult to supply a universal quantitative expression. 

We estimate $\kappa_{xx}^\mathrm{c}$ using typical parameters
$\kappa_{xx}^{\mathrm{e}} = 1000~\mathrm{W/m~K}$ and $\kappa_{xx}^{\mathrm{p}} = 0.2~\mathrm{W/m~K}$ at T = 20 K~\cite{smith1995low, AMILS201623}, corresponding to electrical conductivities of $2 \times 10^{9}~\mathrm{S/m}$ and $4 \times 10^{6}~\mathrm{S/m}$, respectively. Setting $t = 10~\mu\mathrm{m}$, $l_{c} = 100~\mu\mathrm{m}$, and $t^{\prime}=1~\mu \mathrm{m}$, we estimate $\kappa_{xx}^\mathrm{c} \approx 50 \mathrm{W/m\ K}$, corresponding to an electrical conductivity of $10^8~\mathrm{S/m}$.
This is the value used in the main text.

Thermal boundary resistance could also influence the effective thermal conductivity.
It can be considered as an additional thin layer besides the silver paint.
This introduces an extra term into Eq.~(\ref{eq.C2}):
\begin{equation}
\frac{1}{\kappa^\mathrm{c}_{xx}} \approx \frac{1}{\kappa^{\mathrm{e}}_{xx}} \frac{t+t'}{t} + \frac{1}{\kappa^{\mathrm{p}}_{xx}}\frac{4 t'(t+t')}{l_{c}^{2}} + R_b\frac{4(t+t')}{l_{c}^{2}},\label{eq.C3}
\end{equation}
where $R_{b}$ denotes the thermal boundary resistivity of the interface between the silver paint and the insulating sample.
The value of $R_{b}$ is not readily available and varies across experiments.
Here, we estimate its contribution by setting $R_b = 2.7 \times 10^{-5}~\mathrm{m^{2}~K/W}$ at $T=20~\mathrm{K}$, based on an extrapolation of data from Ref.~\onlinecite{schmidt1974thermal} for the sapphire–epoxy interface.
This gives $\kappa^\mathrm{c}_{xx} \approx 11~\mathrm{W/m~K}$, which is roughly 1/5 of the value used in the main text.
It means that the nominal $W/t$ values obtained in the main text should be scaled by the same factor in this case.

%Any resulting inaccuracy can be effectively absorbed into a scaling of the nominal parameter $W/t$.

\bibliography{apssamp}% Produces the bibliography via BibTeX.

\end{document}